\documentstyle[emulateapj,pstricks]{article}

\def\Mvir{M$_{vir}${\ }}
\def\rvir{$r_{vir}${\ }}
\def\LCDM{$\Lambda$CDM{\ }}
\def\LCDMs{$\Lambda$CDM$_{7.5}${\ }}
\def\LCDMl{$\Lambda$CDM$_{15}${\ }}

\begin{document}
\submitted{to be published in July 20, 1998 issue of the Astrophysical Journal}
\slugcomment{{\em to be published in July 20, 1998 issue of the Astrophysical Journal}}
\lefthead{CORES OF DARK MATTER-DOMINATED GALAXIES}
\righthead{KRAVTSOV ET AL.}

\title {THE CORES OF DARK MATTER-DOMINATED GALAXIES:\\ THEORY VERSUS
  OBSERVATIONS} 
\vspace{1mm}
\author{Andrey V. Kravtsov and Anatoly A. Klypin}
\affil {Astronomy Department, New Mexico State University, Las Cruces,
  NM 88003-0001, USA} 
\and \author {James S. Bullock and Joel R.
  Primack} 
\affil {Physics Department, University of California, Santa
  Cruz, CA 95064}
%
\begin{abstract}
  
  We use the rotation curves of a sample of dark matter dominated dwarf
  and low-surface brightness (LSB) late-type galaxies to study their
  radial mass distributions. We find that the shape of the rotation
  curves is remarkably similar for all (both dwarf and LSB) galaxies in
  the sample, suggesting a self-similar density distribution of their
  dark matter (DM) halos. This shape can be reproduced well by a
  density profile with a shallow central cusp $[\rho(r)\propto
  1/r^{\gamma},{\ }\gamma\approx 0.2-0.4]$ corresponding to a steeply
  rising velocity curve $[v(r)\propto r^g,{\ } g\approx 0.9-0.8]$.  We
  further show that the observed shapes of the rotation curves are well
  matched by the average density profiles of dark matter halos formed
  in very high resolution simulations of the standard cold dark matter
  model (CDM), the low-density CDM model with cosmological constant
  ($\Lambda$CDM), and the cold$+$hot dark matter model with two types
  of neutrino (CHDM).  This is surprising in light of several previous
  studies, which suggested that the structure of simulated dark matter
  halos is inconsistent with the dynamics of dwarf galaxies. We discuss
  possible explanations for this discrepancy and show that it is most
  likely due to the systematic differences at small radii between the
  analytic model proposed by Navarro, Frenk, \& White, with
  $\gamma_{\rm NFW} = 1$, and the actual central density profiles of
  the dark matter halos. We also show that the mass distributions in the
  hierarchically formed halos are {\em on average} consistent with the
  shape of rotation curves of dark matter dominated galaxies. However,
  the scatter of the individual profiles around the average is
  substantial.  Finally, we show that the dark matter halos in our
  hierarchical simulations and the real galaxies in our sample exhibit
  a similar decrease in their characteristic densities with increasing
  characteristic radial scales and show increase in their maximum rotation
  velocities with increase in the radii at which their maximum velocities
  occur.
  
\end{abstract} 
\keywords{cosmology: theory -- dark matter: halos --- galaxies:
  kinematics and dynamics --- galaxies: structure}
%
\section{INTRODUCTION}
The amount of luminous matter (stars and gas) in many 
spiral and irregular galaxies is not sufficient to explain the amplitude
and shape of their rotation curves (RCs). This discrepancy 
is usually interpreted as evidence 
for the presence of an extended dark matter (DM) halo surrounding the visible 
regions of galaxies (e.g., Casertano \& van Gorkom 1991;
Persic, Salucci, \& Stel 1996, and
references therein). The extent of the dark matter halos, 
estimated using satellite dynamics, is\footnote{We assume 
that the present-day value of the Hubble 
constant is $H_0=100h$ km/s/Mpc.}$\sim 0.2-0.5h^{-1}$ Mpc
(Zaritsky \& White 1994; Carignan et al. 1997; Zaritsky et al. 1997). 
However, the
dynamical contribution of the dark matter can be substantial even in
the very inner regions of galaxies:
the observed rotation velocities of some dwarf and low-surface brightness
(LSB) galaxies imply that DM constitutes a dominant 
fraction (up to $\sim 95\%$) 
of dynamical mass within the last measured point of their RCs
(e.g., Carignan \& Freeman 1988; Martimbeau, Carignan, \& Roy 1994; 
de Blok \& McGaugh 1997). These {\em dark matter dominated galaxies} offer a 
unique opportunity for probing {\em directly} the density structure of DM halos
which can be then compared with predictions of theoretical models. 

The detailed structure of DM halos formed via dissipationless hierarchical 
collapse in CDM-like models was recently studied using high-resolution 
$N$-body simulations (Dubinski \& Carlberg 1991; 
Navarro, Frenk, \& White 1996, 1997, hereafter NFW96 and NFW97). 
The halo density profiles were found to be cuspy (coreless) and well
fitted by the following two-parameter profile (NFW96):
\begin{equation}
\rho_{\rm{NFW}}(r) = \frac{\rho_s}{(r/r_s)\left(1+r/r_s\right)^2}.
\end{equation}
The characteristic density, $\rho_s$, and radius, $r_s$, are sensitive to 
the epoch of halo formation and are tightly correlated with the 
halo virial mass 
(NFW 96,97). Therefore, the results of these simulations suggest a
coreless and self-similar density structure of   
DM halos, with the virial mass being the single scaling parameter. 

The structure of the inner regions of galactic halos was studied by 
Flores \& Primack (1994) and Moore (1994), who used high-resolution
rotation curve measurements of 
several dark matter dominated dwarf galaxies. The  
central density distributions in these galaxies were found to be 
inconsistent with the singular 
[$\rho(r)\propto 1/r$] behavior predicted by equation (1). The scaling 
properties of the observed halos 
were analyzed by Burkert (1995, 
hereafter B95), who pointed out that shapes of the density profiles
of four dwarf galaxies analyzed by Moore (1994) are remarkably similar 
and are well fitted by the following phenomenological density profile:
\begin{equation}
\rho_{\rm{B}}(r) = \frac{\rho_b}{(1+r/r_b)\left[1+(r/r_b)^2\right]}.
\end{equation}
Parameters $\rho_b$ and $r_b$ were found to be strongly correlated, in 
qualitative agreement with the predictions of hierarchical models (B95). 

In this paper we study the observed density structure in a sample of 
dark matter dominated galaxies inferred from their rotation curves. 
Particularly, we test two predictions of previous simulations of hierarchical
halo formation: (1) cuspy central density distribution and 
(2) self-similarity of the halo density structure. 
We then use results of high-resolution $N$-body 
simulations to compare the observed rotation curves with 
circular velocity profiles of dark matter halos formed in different 
structure formation models. 
\begin{deluxetable}{lcccccccccc}
\tablewidth{0pt}
 \tablecaption{The sample of dwarf and LSB galaxies}
\tablehead{
\colhead{  }&\colhead{ }&\colhead{$r_0$}&\colhead{$V_0$}&\colhead{$\rho_0$}&\colhead{$r_t$}&\colhead{$V_t$}&\colhead{$r_{max}$}&\colhead{$V_{max}$}&
Distance&  \nl 
Galaxy    & $M_B$&$h^{-1} {\rm kpc}$&km s$^{-1}$&$10^8h^3
M_{\odot} {\rm kpc^{-3}}$&$h^{-1} {\rm kpc}$&km s$^{-1}$&$h^{-1}
  {\rm kpc}$&km s$^{-1}$&$h^{-1} {\rm Mpc}$&Reference\nl
(1) &(2) &(3) &(4) &(5) &(6)&(7)&(8)&(9)&(10)&(11) \nl
}
\startdata
        & & & & &   Dwarf           & & & & &    \nl
 DDO 154  &-13.8&2.0&38&0.33&2.9& 76& 5.6& 47&3.0& 1,2 \nl
 DDO 170  &-15.2&4.4&52&0.13&6.5&105&12.4& 64&14& 3   \nl
 NGC 2915 &-16.8&1.1&69&3.77&1.7&140& 3.2& 86&2.5& 4   \nl
 IC 2574  &-15.7&5.1&65&0.15&8.2&138&15.6& 85&2.0& 5   \nl
 NGC 5585 &-17.5&2.3&73&0.93&3.4&148& 6.5& 91&4.7& 6   \nl
 DDO 236  &-16.8&4.4&59&0.16&6.8&122&13.1& 75&1.4& 7   \nl
 DDO 7    &-17.7&3.8&87&0.49&5.5&176&10.5&108&18& 8   \nl
 DDO 10   &-16.3&3.5&54&0.23&5.3&112&10.0& 69&7.8& 8   \nl
 UGC 2684 &-13.7&1.6&41&0.59&2.6& 86& 4.9& 53&4.1& 8   \nl
 DDO 34   &-15.7&1.5&54&1.29&2.1&109& 4.0& 67&5.9& 8   \nl 
          &     &   &  &    & LSB  &   &    &   &   &     \nl
 F568-1   &-17.5&3.8&97&0.61& 5.5&197&10.5&121&64& 9   \nl
 F568-3   &-17.7&6.1&96&0.23& 9.1&196&17.4&121&58& 9   \nl
 F568-V1  &-17.3&4.9&96&0.36& 7.2&194&13.7&119&60& 9   \nl
 F571-8   &-17.0&6.8&121&0.30&10.1&248&19.4&153&36& 9   \nl
 F574-1   &-17.8&8.8&95&0.11&13.6&197&25.9&121&72 & 9   \nl
 F583-1   &-15.9&4.6&71&0.22& 6.7&143&12.8& 88&24& 9   \nl
 F583-4   &-16.3&6.0&64&0.11& 9.3&133&17.8& 82&37& 9   \nl
\tablenotetext{ } 
{NOTES.-- Col.(2) $M_B$, 
blue absolute magnitude; col.(3) best fit $r_0$(see eq. [3]; 
the fitting procedure is described in \S2.2);
col.(4) $V_0=V(r_0)$; 
col.(5) best fit $\rho_0$ (see eq.[3]);
col.(6) best fit $r_t$ (see eq.[4]); col.(7) best fit $V_t$ 
 (see eq.[4]); col.(8) $r_{max}$; 
  col.(9) $V_{max}=V(r_{max})$; col.(10) distance to galaxy adopted 
   in this study;
 }
\tablenotetext{ }
 { REFERENCES.--
  (1) Carignan \& Freeman 1988; 
  (2) Carignan \& Beaulieu 1989; 
  (3) Lake et al. 1990;
  (4) Meurer et al. 1996;
  (5) Martimbeau et al. 1994;
  (6) C\^ot\'e et al. 1991;
  (7) Jobin M. \& Carignan C. 1990;
  (8) van Zee et al. 1997;
  (9) de Blok et al. 1996. 
 } 
\enddata
\end{deluxetable}
%
\section{DWARF AND LSB GALAXIES}
\subsection{{\em The sample}}
We have compiled a sample of 10 dwarf and 7 LSB galaxies 
with measured rotation curves and published mass models for stellar, gas, 
and halo components (see Table 1). The dwarf galaxies were selected 
from different sources, whereas all 7 LSB galaxies were selected from 
the sample of de Blok, McGaugh, \& van der Hulst (1996) (see Table 1).  

In our sample we included only those galaxies in which the dark matter 
component was shown to constitute $\gtrsim 85$\% of the 
total mass inside the last measured point of the rotation curve
(in most cases with the maximum disk assumption). 
It is important to note that distances to all of the dwarf, and some of 
the nearby LSB, galaxies are quite uncertain (in some cases by a factor 
of two). While rotational velocity is a directly observable quantity, 
the physical scale of rotation curves must be computed from the angular 
scale using distance. Thus, any uncertainty in the distance propagates 
into uncertainty in the physical scale. This fact should be kept in mind 
when one makes a one-to-one comparison of observed and modeled 
rotation curves. The distances to the galaxies adopted in our analysis 
are listed in Table 1. We have adopted the best estimate of
distance from the original paper, when it was available, or the distance 
quoted in Tully (1988). We have also included in Table 1
the best fit values of mass model parameters described in \S 2.2. 
%
\subsection{{\em Rotation curve analysis}}
%
Analysis of the dark matter distribution is difficult in most 
galaxies due to ambiguities in the estimates
of the stellar mass-to-light ($M/L$) ratios and the resulting 
dynamical contribution
of the stellar component to the observed rotation velocities 
(e.g., de Blok \& McGaugh 1997; Bottema 1997; Courteau \& Rix 1997; and 
references therein). However, the rotation curves of the galaxies in our 
sample are mostly determined by dark matter on scales $\gtrsim 1$ kpc:
the contribution from the gas and stars 
is negligible and residuals between the observed rotation curves and
contribution of DM are at the level of the observational scatter of 
rotational velocities.
Therefore, the dark matter distribution models can be directly fitted 
to the rotation curves of these galaxies without uncertain
assumptions about $M/L$ ratios. 

B95 showed that the density distribution, $\rho_B(r)$,
described by equation (2)
fits the data very well over the entire observed range of scales. At large 
radii this profile falls off as $\rho(r)\propto r^{-3}$, in accord with 
simulations of the CDM model (e.g., NFW96). However, the change of 
logarithmic slope at $r\sim r_0$ predicted by $\rho_{\rm{B}}(r)$, 
equation (2), is faster
than the change predicted by $\rho_{\rm{NFW}}(r)$,
equation (1). 
Moreover, $\rho_{\rm{B}}(r)$ has 
a flat core at small radii ($r \ll r_{\rm b}$), in disagreement with
the $r^{-1}$ central cusp of $\rho_{\rm{NFW}}(r)$. 
For dwarf galaxies, the scale at which the 
density distribution is expected to become flat is quite small 
 ($\lesssim 1$ kpc) and is in fact below the current 
observational resolution. From a theoretical point of view the existence of a
core is difficult to understand because hierarchical formation of halos is 
much more likely to result in cuspy central density distributions 
(Syer \& White 1997). 
Therefore, we will consider the broader family of density profiles 
(Zhao 1996):
\begin{equation}
\rho (r)=
\frac{\rho_0}{(r/r_0)^{\gamma}[1+(r/r_0)^{\alpha}]^{(\beta-\gamma)/\alpha}}.
\end{equation}
Note that $\rho(r \ll r_0) \propto r^{-\gamma}$, $\rho(r \gg r_0) \propto
 r^{-\beta}$, and $\alpha$ characterizes the sharpness of the 
change in logarithmic slope.
This family includes both cuspy profiles of the type proposed by NFW96 
$(\alpha, \beta, \gamma)=(1,3,1)$ and the so-called modified isothermal 
profile $(\alpha, \beta, \gamma)=(2,2,0)$, which is the most widely used 
model for the halo density distribution in analyses of observed 
rotation curves. It is also convenient to make direct fits
with an analytic model similar to (3) for the velocity profile:
\begin{equation}
V(r)=V_t\frac{(r/r_t)^g}{[1+(r/r_t)^a]^{(g+b)/a}},
\end{equation}
where $r_t$ and $V_t$ are the effective ``turnover'' radius and
velocity and $a$ parameterizes the sharpness of the turnover.  
The limiting behaviors are $V(r \gg r_t) \propto 1/r^{b}$
and $V(r \ll r_t) \propto r^g$.  The peak of the velocity profile (4)
occurs at the radius $r_{\rm max} = r_t(g/b)^{1/a}$, and
$V_{max} = V(r_{max}) = V_t(g/b)^{g/a} [1+g/b]^{-(b+g)/a}$. 

The existing rotation curve measurements,
due to their finite resolution and extent,
cannot be used to constrain all five parameters of the profiles (3) and
(4).  The inner radii of the observational rotation curves are well probed, 
so we can make a meaningful comparison with fitting functions
having different inner density profiles $\gamma$. 
However, most of the galaxies in our sample have rotation
curves that have not begun to decline at 
the outermost measured point and thus have
very little information about the asymptotic slopes $\beta$, $b$.
Given the uncertainties, we fix the outer logarithmic slope to the
value suggested by the models\footnote{
$V_{\rm NFW}(r) \propto \sqrt{\ln r/r}$ for large $r$, which has
an approximate slope of $b\sim0.34$ for values of r near a typical virial
radius.} (1) and (2): $\beta=3$, $b=0.34$. Note, however, that as was
noted by Burkert (1995) this value is also favored (to $\beta=2$ of
the pseudo-isothermal profile) by {\em observed} highest
quality rotation curves of dwarf galaxies.
 For the
same reason, the sharpness of the turnover, $\alpha$ or $a$, 
are not constrained for all of the galaxies. 
However, a fair number of galaxies in the sample do
 show the turnover and thus can be used to constrain $\alpha$.
 The plausible value of the parameter $\alpha=2$ was determined using
 rotation curves of these galaxies.
 We generalize this value to all of the galaxies (which in no way
 contradicts the data, but is not, of course, a strict procedure) and thus
 currently we can only talk about a
 plausible range of $\alpha$ values as a ``universal fit'' (if any such
 universal value exist at all). For example, our results will not 
 change drastically if we use $\alpha=1.5$ instead of
 $\alpha=2$. However, $\alpha=1$ gives a poorer fit to the data. 
 
We fix the parameter $\gamma$ to $0.2$: 
the value which 
best fits most of the observed rotation curves. 

The corresponding best-fit slopes of the profile (4) are
$(a,b,g)=(1.50,0.34,0.9)$.  Note that $g=1-\gamma/2$.
With parameters $\alpha$, $\beta$, and $\gamma$ $(a,b,g)$ fixed, 
we fitted the data for the remaining free parameters of the profile (3): 
$\rho_0$ and $r_0$ ($V_t$ and $r_t$ in eq.[4]). Our fits thus have the same 
number of free parameters as do profiles (1) and (2). 
Note that while the particular set of the parameters 
 ($\alpha$, $\beta$, $\gamma$)$=(2,3,0.2)$ used in the paper didn't
 result from a strict fitting procedure, it was motivated by all
 possible constraints of available data. The only theoretically
 suggested value is that of $\beta$ but it also seems to be favored by
 data (Burkert 1995). Hopefully, as new RC observations come along, 
 they can be used to 
 pinpoint the parameters $\alpha$ and $\beta$ with a better accuracy.

\pspicture(0.5,-1.5)(13.0,10.6)

\rput[tl]{0}(-0.5,11.3){\epsfxsize=10cm
\epsffile{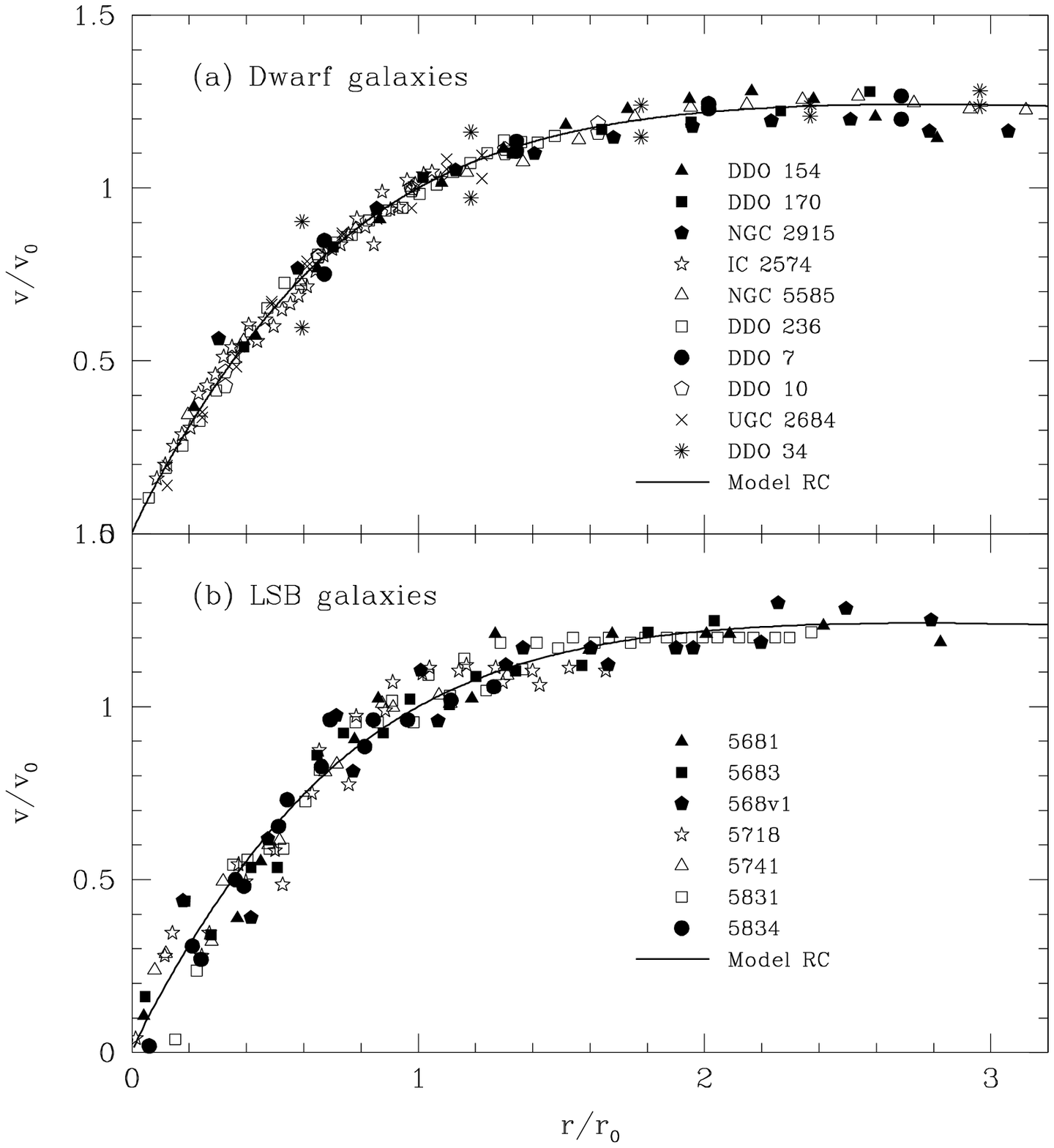}}

\rput[tl]{0}(0,1.8){
\begin{minipage}{8.7cm}
  \small\parindent=3.5mm {\sc Fig.}~1.---Rotation curves of (a) dwarf
  and (b) LSB galaxies (symbols) normalized to the best fit values of
  $r_0$ and rotational velocities $v_0$ at $r_0$ predicted by density
  model (eq.[3]).  The solid line on both panels shows the analytic
  rotation curve corresponding to the density profile (eq.[3]) with
  $(\alpha, \beta, \gamma)=(2,3,0.2)$.  The rotation curves for
  different dwarf {\em and} LSB galaxies have virtually identical
  shapes, which is very well matched over the entire observed range of
  scales by the analytic model. Note, that the RC of NGC 2915 extends
  outside the scale of the plot: the outer part of this RC can be seen
  in Figure 2.
\end{minipage}
}
\endpspicture

Figure 1 shows rotation curves of
dwarf (a) and LSB (b) galaxies normalized to their best 
fit values of $r_0$ and to the rotational velocities $v_0$ at $r_0$, 
predicted by analytic profile (3). The best fit values of $r_0$,
$\rho_0$, $V_0=V(r_0)$, $r_t$, and $V_t$ are given in Table 1 for each
galaxy in the sample. The formal errors of each of these values are
less than about $\sim 2-5\%$.  
Figure 1a shows that all of the dwarf galaxies have rotation curves of 
virtually identical shape\footnote{By the {\em shape} 
of a rotation curve we mean its particular functional form. For example,
the shape of the rotation curve described by equation (4) is $x^g/(1+x^a)^{(g+b)/a}$ (where $x\equiv r/r_t$). By saying that the RC shape is similar for all
our galaxies, we mean that all their rotation curves
 can be described by this functional form 
with fixed values of parameters $a$, $b$, and $g$.}
 with a remarkably small scatter.
The rotation curves of the two dwarf galaxies, DDO154 and NGC2915, 
cannot be described by a smooth density distribution model 
in their outer parts. The RC of DDO154 shows a decrease in rotational velocity
in the three outermost observed points. Conversely, the RC of NGC2915 has a sharp
upturn at $\gtrsim 5h^{-1}$ kpc (or $r/r_0\gtrsim 4.5$). This upturn
can be seen in Figure 2. The explanation of this peculiar behavior
is not clear (see, however, Burkert \& Silk 1997), but it is obvious
that it cannot be explained by any smooth model for the mass distribution.
Note, however, that apart from the peculiar outer regions, 
the rotation curves of both DDO154 and NGC2915 have the same shape as the rest
of the galaxies. 

The shape of the galaxies' rotation curves is well matched by the rotation 
curve corresponding to 
density profile (3) with $(\alpha, \beta, \gamma)=(2,3,0.2)$ or 
correspondingly to RC (4) with $(a,b,g)=(1.5,0.34,0.9)$. This
result is in perfect agreement with Burkert (1995) who showed
similarity of rotation curves for four dwarf galaxies
(two of which, DDO154 and DDO170, were included in our sample). 
As was mentioned above, the similar fit by $\rho_{\rm B}(r)$
(eq. [2]) proposed by Burkert (1995) is equally good. 
Note, however, that our profile does not have any flat core, whereas
$\rho_{\rm B}(r)$ predicts such a core at $r\ll r_b$. The fact that both 
profiles fit the data equally well is easy to understand if we 
notice that $\rho_{\rm B}(r)$ predicts a flat density distribution at the 
scales well below the observational resolution ($\lesssim 1$ kpc). 
 Thus, $\rho_{\rm B}(r)$ and profile (3) can be virtually identical
in the range of scales resolved in observations and thus provide 
an equally good fit to the data.

Figure 1b shows that the rotation curves of dark matter dominated
LSB galaxies are also well
described by the {\em same} analytic density profile. 
The larger amplitude of scatter in the case of LSB galaxies 
can be explained by the larger observational errors associated with a given
point of a rotation curve and thus most likely reflects observational
uncertainties rather than intrinsic scatter of the
halo properties. Most of the LSB galaxies in our sample are located
at considerably larger distances than dwarf 
galaxies. Therefore, the dwarf galaxies have been 
observed with considerably higher resolution and smaller observational errors 
than LSB galaxies. The estimated errors are typically $10-20\%$ (de Blok et
al. 1996), especially in the 
inner regions of galaxies ($\lesssim 10$ kpc).  

One important issue is whether subtraction of baryon component (stars
and gas) in the galaxies from our sample will affect results of the
rotation curve analysis presented above. As we mentioned in \S 2.1,
the combined contribution
  of stars and gas is ${\lesssim} 15\%$ for all of our galaxies 
  (${\lesssim} 10\%$ in most cases). It is clear that ideally one has
  to subtract contributions of both gas and stars from the observed RC
  in order to get the mass distribution of the dark matter to an
  accuracy of better than 10\%. However, it
  is well known that this is not an easy thing to do. For stars, 
  the exact mass-to-light ratio is not known and we cannot convert
  visible flux into the stellar mass without making additional
  assumptions (hence, ``maximum disk controversy''). In the case of
  gas, we know exactly how to convert the 21-cm flux into the mass
  of gas (no mass-to-light uncertainty). This conversion, however, relies on
  other assumptions which can easily lead to uncertainties as high as
  10\% in the contribution of gas. For example, we need to know distance to
  the galaxy in order to make this conversion. We also need to have a
  reliable way to estimate the profile of molecular gas to recover the
  dark matter density profile. The distances to the
  dwarf galaxies in our sample are {\em very} uncertain (often by a
  factor of two or more) and so is the conversion. It is not clear
  whether it makes any sense to subtract the gas with
  uncertainties of its contribution this big. 
  After all, the  motivation for use of
  dark matter dominated galaxies for this kind of analysis is to avoid
  dubious or uncertain correction procedures 
  which are very unlikely to result in a better
  determination of the shape of DM density distribution.  
  The distance uncertainty is not so severe for LSB galaxies and so
  conversion could, in principle, have been done in this case. We
  have not done this for one simple reason: subtraction of the gas
  component could change any given point of rotation curve by at most 
  $10\%$ (in practice less than that). However, the errors
  associated with each point of RC are of the order of $10-20\%$ (which
  combine both observational errors and assymetries in the rotation
  curves between receding and approaching sides) and 
  it seems unlikely that correction due to gas subtraction would improve
  or systematically change the answer (unless the observational errors
  are significantly overestimated). 
  We  have tested the effect of gas
  subtraction on the RC shapes by performing RC shape analysis for 
  two galaxies (dwarf NGC 5585 and LSB F$583-1$) with and without 
  subtraction of gas. The two galaxies have been selected to have a
  clearly visible turnover of the RC and to have a fairly high fraction
  of gas inside the last measured point of the rotation curve. This
  fraction is 8\% for NGC 5585 (C\^ote et al. 1991) 
  and 5\% for F$583-1$ (de Blok et al. 1996). The results of
  fitting the ($\alpha$, $\beta$, $\gamma$)$=(2.0, 3.0, 0.2)$ model to
  the RC with and without gas subtraction result in very similar
  results: the difference in the best fit parameters is $\lesssim 10\%$
  and corrected and uncorrected RCs have virtually identical shape. 
  Note also that the dwarf galaxies that we used have on average a higher 
  (or at least as high)
  fraction of gas (typically $\sim 6-10\%$) than LSB galaxies
  (typically $3-7\%$). Therefore, if gas would introduce systematic
  differences in the shape of RC, we could expect that the scatter in
  Figure 1 to be larger for dwarf galaxies (some
  galaxies have much more gas than the others). Yet, the shape of
  rotation curves for dwarf galaxies is very uniform.

We have repeated the fitting procedure described above using the analytic 
profiles (1) and (2). As was mentioned above, $\rho_{\rm B}(r)$ 
results in a fit that is equally good
to the fit by profile (3) shown in Figure 1. However, the analytic 
profile proposed by NFW failed to produce a reasonable fit to
the data, as was indeed pointed out in NFW96 (see their fig.12).
The major difficulty with this profile, as was noted before
by Flores \& Primack 1994 and B95, is that the inner slope of the density 
distribution ($\gamma=1$) is considerably steeper than implied by the 
rotation curves. The finite spatial extent of the data and incorrect
inner slope of the profile (1) lead to implausible solutions of the 
$\chi^2$-minimization 
procedure (the values of $r_s$ increase without convergence). 

The observed similarity of the shapes of the rotation curves for
seventeen different galaxies, selected solely on the basis of their
dark matter content, and the remarkably small amount of scatter,
implies that their matter distributions are {\em self-similar} in
terms of the density structure.  Of course, this includes both stellar
and gaseous matter as well as DM.  Both stellar and gaseous masses are
uncertain because of uncertainties in the distance, mass-to-light
ratio, and atomic-to-molecular gas ratio of each galaxy.  To the
extent that we can neglect the stellar and gaseous components (a
subject that we intend to address in a subsequent paper), the
self-similar rotation curves of these DM-dominated galaxies imply that
they all have the same density structure.  The question we now ask is
whether the disagreement between this density structure and $\rho_{\rm
NFW}(r)$ indicates a failure of CDM-type models?

%
%
%
\section{COMPARISON WITH THEORETICAL MODELS}
%
\subsection{{\em Numerical simulations}}

We have used the new Adaptive Refinement Tree (ART) $N$-body code 
(see Kravtsov, Klypin, \& Khokhlov 1997 for details) to simulate 
the evolution of 
collisionless dark matter in the three cosmological structure formation
models: 
(1) standard cold dark matter model 
(CDM: $\Omega_0=1$, $h=0.5$, $\sigma_8=0.7$);
(2) a low-density CDM model with cosmological constant
($\Lambda$CDM: $\Omega_{\Lambda}=0.7$, $h=0.7$, $\sigma_8=1.0$);
and (3) a cold$+$hot dark matter model with two types
of neutrino (CHDM; $\Omega_0=1$ and $\Omega_{\nu}=0.2$; $h=0.5$;
$\sigma_8=0.7$; cf. Primack et al. 1995).
Here $\Omega_0$, $\Omega_{\Lambda}$, and $\Omega_{\nu}$ are the
present-epoch values of the density of matter, 
vacuum energy (as measured by the cosmological constant), 
and massive neutrinos, respectively. The rms fluctuation in spheres of 
radius $8h^{-1}$ Mpc, $\sigma_8$, was chosen to conform with the local 
abundance of galaxy clusters, for $\Lambda$CDM and CHDM models
it is also in agreement with 
measurements of the cosmic microwave background anisotropy by the {\sl
COBE} satellite.
 The simulations followed trajectories of $128^3$ cold 
dark matter particles in a box of size of $L_{box}=7.5h^{-1}$ Mpc. 
In the CHDM simulation, two additional equal-mass ``massive neutrino'' 
species were evolved, which brings the number of particles in the simulation
to $3\times 128^3$. 
To test for the possible effects of the
finite box size, we have run an additional simulation of the \LCDM  model
with the box size twice as large: $L_{box}=15h^{-1}$ Mpc$=21.43$ Mpc. 
We will denote the two \LCDM  simulations as \LCDMs and \LCDMl
according to their box sizes. 

We have used a $256^3$ uniform 
grid covering the entire computational volume and finer refinement
meshes constructed recursively and adaptively inside the high-density regions.
The {\em comoving} cell size corresponding to a refinement level $L$ is
$\Delta x_L=\Delta x_0/2^L$, where $\Delta x_0=L_{box}/256$ is the size of
the uniform grid cell ($L=0$ corresponds to the uniform grid). 
The increase of spatial resolution
corresponding to each successive refinement level was accompanied by
the decrease of the integration time step by a factor of 2. 
The simulations were started at redshift $z_i=40$ in the CDM and \LCDMs
simulations and at $z_i=30$ in the CHDM and \LCDMl simulations. 
Particle trajectories 
were integrated with the time step of $\Delta a_0=0.0015$ on the 
zeroth-level uniform grid in the case of the CDM and $\Lambda$CDM runs
and with $\Delta a_0=0.006$ in the CHDM run. The time 
step on a refinement level $L$ is $\Delta a_L=\Delta a_0/2^L$.
The time step for the highest refinement level corresponds to 
$\gtrsim 40,000$ time steps over the Hubble time.    
Six refinement levels were introduced in the highest density regions 
corresponding to a cell size of $\Delta x_6=0.46h^{-1}$ kpc. The
{\sl dynamic range} of the simulations is thus $256\times 2^6=16,384$. 
Note, that the resolution is {\em constant in comoving coordinates} which 
means that actual physical resolution is higher at earlier epochs (the halos
were resolved with six refinement levels as early as $z\approx1$).
The
refinement criterion was based on the local overdensity of dark matter 
particles. Regions with overdensity higher than 
$\delta = n_{th}(L){\ }2^{3(L+1)}$ were refined to the refinement level
$L$. Here, $n_{th}(L)$ is the 
threshold number of particles per mesh cell of level $L$ estimated using the 
cloud-in-cell method (Hockney \& Eastwood 1981).  
We have used values $n_{th}=5$ at all levels in the CDM and
$\Lambda$CDM runs; for the CHDM run we have used $n_{th}=10$ at the levels
$L=0,1$ and $n_{th}=5$ for all of the higher levels. These values of threshold
were suggested by results of the tests presented in Kravtsov et
al. (1997); they ensure that refinements are introduced only in the
regions of high-particle density,
where the two-body relaxation effects are not important. 

For the dark matter halos used in our analysis the spatial resolution
is equal to $\approx 0.5-2h^{-1}$ kpc (corresponding to the 6th to 4th 
refinement levels). For each of the analyzed halos, we have taken
into account only those regions of the density and circular velocity 
profiles that correspond to 
scales at least twice as large as the formal resolution.  
The mass resolutions (particle mass) of our simulations
 are listed in Table 2, and are in the range of 
$\sim (1-10)\times 10^7h^{-1}$ M$_{\sun}$. Therefore a typical halo 
of mass $\sim 10^{11}h^{-1}$ M$_{\sun}$ in our simulations contains
several thousands of particles. 
  
These simulations are comparable in spatial and mass resolution, as well
as in the box size, to those of NFW96,97. There is, however, a
significant difference: our simulations are direct simulations of 
{\em all} DM halos in a given computational volume, whereas 
NFW96,97 simulate with high resolution a handful of individual halos. 
The fact that we analyze a statistically large sample consisting of
dozens of galaxy-size halos in each simulation allows us to make 
conclusions about {\em average} halo properties and estimate the amount
of cosmic scatter. A summary of the numerical simulations is given in 
Table 2. The parameters listed in this table are defined in the 
text above. 
\begin{deluxetable}{lccccccccc}
\scriptsize
\tablewidth{0pt}
 \tablecaption{Parameters of the Numerical Simulations}
\tablehead{
\colhead{Model}&\colhead{$\Omega_0$}&\colhead{$\Omega_{\Lambda}$}&
\colhead{$\Omega_{\nu}$}&
\colhead{$h$}  &\colhead{$\sigma_8$}&\colhead{$z_i$}             &
\colhead{$\Delta a_0$}&\colhead{$L_{box}$}&\colhead{Particle mass}\nl
               &          &          &                            &
               &          &          &   $\times 10^{-3}$         &
\colhead{$h^{-1}$ Mpc}& \colhead{$\times 10^7h^{-1}$ M$_{\odot}$}          
}
\startdata
CDM                 &1.0 & 0.0 & 0.0 & 0.5 & 0.7 & 40 & 1.5 & 7.5 & 5.56 \nl
$\Lambda$CDM$_{7.5}$&0.3 & 0.7 & 0.0 & 0.7 & 1.0 & 40 & 1.5 & 7.5 & 1.67 \nl
$\Lambda$CDM$_{15}$ &0.3 & 0.7 & 0.0 & 0.7 & 1.0 & 30 & 1.5 &15.0 & 13.3 \nl
CHDM                &1.0 & 0.0 & 0.2 & 0.5 & 0.7 & 30 & 6.0 & 7.5 &
4.44\tablenotemark{a}, 0.56\tablenotemark{b}\nl
\tablenotetext{a}{cold particles; $^b$ hot particles.}                  
\enddata
\end{deluxetable}
%
\subsection{{\em Tests of numerical effects}}
%

There are several effects which can affect the halo density profiles
at scales larger than some particular scale related to this effect. 
We have tested the reliability of the simulated density and velocity profiles 
by comparing results of the simulations with different
resolutions and time steps. Specifically, the tests were used to
determine the range of numerical parameters for which the convergence of
density profiles was found at scales larger than two formal resolution
elements (formal resolution is equal to the size of the refinement mesh
cell).
 
Tests presented in 
Kravtsov et al. (1997) show that the density profiles are not affected
by the force resolution down to a scale of about one resolution element
(a similar conclusion was reached by NFW96). To test the effects of the
time step we have used a set of $64^3$-particle simulations of the CDM 
model with parameters identical to those described in the previous
section. These test simulations were started from identical initial
conditions, but evolved with different time steps: 
$\Delta a_0=0.006,0.003,0.0015,0.00075$.
Comparison of the density profiles for the {\em same} halos in these 
simulations shows that for halos of all masses, the profiles converge
for runs with  $\Delta a_0\lesssim 0.0015$ (the value used in our 
CDM and $\Lambda$CDM simulations) at all scales, down to the 
resolution limit. We further use two $128^3$-particle simulations of 
the \LCDM model with the box size of 
$15h^{-1}$ Mpc and with time steps of $\Delta a_0=0.006$ 
and $\Delta a_0=0.0015$.
The comparison shows that the most massive halos 
(virial mass \Mvir$>10^{13}h^{-1}$ M$_{\sun}$) have systematically shallower 
central ($r\lesssim 10-20h^{-1}$ kpc) density profiles 
in the $\Delta a_0=0.006$ run as compared to the halos from 
the $\Delta a_0=0.0015$ run. 
However, the difference decreases with decreasing halo mass and for 
\Mvir$\lesssim 5\times 10^{12}h^{-1}$ M$_{\sun}$ the density profiles
from the two runs are identical within statistical noise at scales
larger than one resolution element. 
This mass dependence is due to the different accuracy of numerical
integration in objects of different masses. The accuracy depends on the 
average displacement of particles during a single time step:
for the integration to be accurate, the displacement should be $\lesssim$
$10-20\%$ of the resolution element. 
Particles inside more massive halos have considerably higher velocities
($v\gtrsim 300-400$ km/s) and thus average displacements that are 
larger than the displacements of
particles inside small halos ($v\lesssim 200$ km/s). 
In this study we 
focus on the mass distribution of the small halos 
($M\lesssim 1\times 10^{12}$ M$_{\sun}$), for which the tests indicate 
convergence of the density profiles for time steps $\Delta a_0\leq
0.006$. The time step of all simulations presented in this paper,
except for the CHDM simulation, is {\em four times smaller} than the above
value (see Table 2). As an additional test, we have compared average
RC shapes for CDM halos in the $7.5h^{-1}$ Mpc box simulation shown in
  Figure 2a and for halos in the same mass range ($\lesssim 1\times
  10^{12}$ M$_{\odot}$) 
from an identical simulation (identical initial
  conditions and simulation parameters) with time step $\Delta
  a_0=0.006$. We have found that average RC shapes and the scatter in these two
  simulations are indistinguishable. 

The mass resolution in our simulations (particle mass) is $\sim(0.6-5)\times
10^7h^{-1}$ M$_{\odot}$ for $L_{box}=7.5h^{-1}$ Mpc runs, and $1.3\times 10^8
h^{-1}$ M$_{\odot}$ for the test $L_{box}=15h^{-1}$ Mpc \LCDM run 
(see Table 2). Therefore, halos of mass $M_{vir}=10^{12}h^{-1}$ M$_{\odot}$
and $M_{vir}=10^{11}h^{-1}$ M$_{\odot}$ (the range of masses used in our comparison with the data) are resolved with $\sim 100,000$ and $\sim 10,000$ particles, respectively. 
For reference, there are $\gtrsim 100-200$ 
particles inside
the innermost point (2 formal resolutions) of the rotation curve 
used in the fitting procedure described below. Comparison of the 
average velocity profiles in the \LCDMs and \LCDMl simulations 
(the latter has {\em eight times} worse mass resolution than the former)
shows that there are no systematic differences between profiles in these
two runs (see Fig.2). 

The force resolution can introduce errors in rotational velocities. 
To estimate this effect, we assume that the finite force resolution
results in a flat core ($\rho=$const) at scales smaller than the 
resolution element $h_r$ in an otherwise ideal NFW halo (Eq. [1]). This results
in the velocity profile $v_{soft}(r)/v_s=\sqrt{f(x)/x f(1)}$, where 
$f(x)\equiv x_h^2/3(1+x_h)^2+F(x)-F(x_h)$, $F(x)\equiv \ln(1+x)-x/(1+x)$,
$x\equiv r/r_s$, $x_h\equiv h_r/r_s$, $v=v(r_s)$, and $r_s$ is the 
scale-radius of the NFW profile (Eq. [1]). This profile can be compared
with the velocity profile corresponding to Eq. (1): 
$v_{NFW}(r)/v_s=\sqrt{F(x)/x F(1)}$. The error is $\sim 18\%$ at 
$r\approx h_r$, and $\lesssim 5\%$ at $r\gtrsim 2h_r$ (see Fig. 5). 
Thus, the velocity profiles of simulated halos should not be significantly
affected at scales $r\gtrsim 2h_r$, which is where we perform the fit to analytic
models.

To test whether the box size of our simulations ($7.5h^{-1}$ Mpc) 
is large enough not to miss all important tidal effects, 
we have compared the density and 
velocity profiles of halos formed in \LCDMs and \LCDMl simulations. 
We have not found any systematic differences between halo 
density profiles in these simulations. 
The average profiles of halos are identical
within the statistical noise (see Figure 2). 
We have also used another indirect way of testing for the proper simulation
of the tidal fields.
Tidal torques from the surrounding large-scale structure presumably play
a major role in the acquisition of the angular momentum, 
$J=\vert{\bf J}\vert$, 
by the galaxy-size halos (Peebles 1969; Doroshkevich 1970). 
Therefore, we can test if the tidal effects were simulated properly 
by comparing the so-called spin parameter for the halos in our
runs with previous results based on the larger-box
simulations. The spin parameter, $\lambda$, of a halo is defined as 
$\lambda\equiv J\vert E\vert^{1/2}/(GM^{5/2})$, where $J$ is the
angular momentum of the halo, $E$ is its total 
energy, and $M$ is the halo virial mass. We have found that the
distributions of $\lambda$ is very nearly log-normal\footnote{Note that
the most probable value of this distribution is $\lambda_{\rm peak}
= \lambda_{*}\exp[-\sigma^2] \sim 0.032$ for the fit values quoted.},
$
        P(\lambda) = (1/\lambda\sqrt{2\pi}\sigma)\exp\left(- 
        {\ln^2(\lambda/\lambda_{*})}/{2\sigma^2}\right),
$
with $\lambda_{*}\approx 0.047$, $0.045$, and
$0.048$ and $\sigma\approx 0.66$, $0.55$, and $0.62$, for the 
CDM, $\Lambda$CDM, and CHDM models respectively.  Our results are
in good agreement with previous studies 
(e.g., Barnes \& Efstathiou 1987; Warren et al. 1992; Cole \& Lacey 1996;
Thomas et al. 1997).
We therefore conclude that our simulations properly include all essential 
tidal effects.   
\subsection{{\em Results}}
We have used a halo-finding algorithm described in Kravtsov et
al. (1997) to identify halos in the simulations at $z=0$. 
The algorithm identifies halos as local maxima of mass inside a given
radius. The exact center of a halo is found iteratively. We have run
tests to ensure that we determine the halo center and 
resulting central density profile correctly. Only halos with more than 
$1000$ particles within their virial radius\footnote{According to 
predictions of the spherical top-hat collapse model
we define the virial radius as radius of a sphere encompassing a mean 
overdensity of 200 for CDM and CHDM models, 
and 340 for $\Lambda$CDM model (Lahav et al. 1991; Kitayama \& Suto
1996; Gross et al. 1997).}, $r_{vir}$, were taken from the full list. 
Also, to avoid effects of ongoing mergers we excluded those halos 
which have close ($r<r_{200}$) companions of mass more than half of the
halo mass. 

The circular velocity profiles, $v(r)=[GM(r)/r]^{1/2}$, were
constructed by estimating the mass inside concentric spherical 
$\Delta =1h^{-1}$ kpc shells around the halo center. 
To avoid contamination by gravitationally unbound background
particles, we iteratively remove all particles which have velocities 
relative to the velocity of the halo as a whole greater than escape 
velocity. The escape velocity, $v_e(r)$, at a given distance $r$ 
from the halo center is computed
analytically assuming that density distribution follows
NFW profile (eq. 1):
$v_e(r)=\sqrt{-2\phi(r)}\approx 2.15{\ }v_{max}\sqrt{\ln(1+2x)/x}$,
where $\phi(r)$ is halo gravitational potential at the distance $r$, 
$v_{max}(r_{max})$
is the maximum rotation velocity of halo and $x\equiv r/r_{max}$. 
This is a good approximation for a large range of 
$r$: NFW96 show that profile (1) approximates the density and velocity 
profiles of halos reasonably well at scales $\sim (0.01-1)$\rvir (see also
Fig.5). The
maximum rotation velocity $v_{max}$ and corresponding scale $r_{max}$
are found from a halo velocity profile at each iteration. The analysis
shows that unbound particles affect at most the outer regions
of halos\footnote{The case, for example, for a small halo located at
the outskirts of a larger system, or for two halos passing close to each
other.}, $r\sim (0.5-1) r_{vir}$, while inner regions are virtually
unaffected (Klypin, Gottl\"ober, \& Kravtsov 1997). 

To compare the shape of the observed and simulated dark matter velocity 
curves, we have fitted the latter with the analytic model described by
equation (3). The parameters $\alpha$, $\beta$, and $\gamma$ were fixed
at the values used to fit dwarf and LSB galaxies -- $2$, $3$, and $0.2$
correspondingly, and we have fitted for the remaining two free parameters --
$r_0$ and $\rho_0$. It should be noted that the observed rotation curves 
of most dwarf and LSB galaxies are measured only to radii of 
$\lesssim 10-30h^{-1}$ kpc and often are still rising at the last
measured point. Therefore, the mass
distribution in the outer parts of the galactic halos (and often
maximum rotation velocity) is poorly constrained. To avoid any bias in
the fitting procedure we considered only the inner $30h^{-1}$ kpc of the
simulated halos. We then normalized each rotation profile to its best fit
values of $r_0$ and rotational velocity $v_0$ at $r_0$ and 
computed the average of these normalized profiles over all halos 
considered in each cosmological model ($\sim 50-60$). In Figure 2 we compare
the average normalized dark matter velocity profiles for halos formed
in CDM, $\Lambda$CDM, and CHDM models, shown by solid lines, 
with corresponding 
profiles of the dwarf galaxies from our sample, shown with different
symbols (the symbols are as in Fig.1). The average profile from the larger-box
\LCDMl simulation is shown with a dashed line. This profile does
not extend to values of $r/r_0$ which are as low as for the \LCDMs profile
(due to worse spatial resolution). However, for values of $r/r_0$,
 where the two profiles overlap, they are indistinguishable. 
  The dotted lines show the
$2\sigma$ envelope representing the scatter of individual halo profiles
around the average. It should be noted that the scatter 
in the inner regions of the halo velocity profiles is substantial.
This scatter possibly reflects physical differences between 
individual halos: our tests show that it is unlikely that 
the scatter can be attributed to 
the statistical noise associated with the finite 
mass resolution. The mass resolution of our simulations
is very high (see Table 2): the number of dark matter particles 
inside the smallest scale, $r_{min}$, of rotation curve used in the fitting 
procedure is $\gtrsim 200$ for large ($\sim 10^{12}$ M$_{\odot}$) halos
and $\gtrsim 100$ for smaller ($\sim 10^{11}$ M$_{\odot}$) halos. 

Figure 2 shows 
that {\em on average} the velocity profiles of halos formed in
hierarchical structure formation models 
 and observed dark matter halos are in good agreement. It also 
shows that both cold dark matter halos and halos of dark matter 
dominated galaxies exhibit a certain self-similarity of the mass 
distribution in their inner regions. It was noted previously (e.g.,
B95, NFW96) that hierarchical formation of the halos should also 
result in well-defined scaling properties of the mass distribution. 
It is thus interesting to compare the scaling properties of 
galaxy halos in our sample with those of the DM halos formed in 
the three hierarchical models studied in this paper. Figure 3 shows
the plot of the best-fit parameters $r_0$ and $\rho_0$ of the model
density distribution (3) for the dwarf (solid circles) and LSB 
(open circles) galaxies together with corresponding parameters 
of DM halos formed in CDM (a), \LCDM (b), and CHDM (c) simulations. 
As before, the values of the remaining parameters of the profile (3)
were fixed to ($\alpha$, $\beta$, $\gamma$)$=$(2,3,0.2). 
For both galaxies and simulated halos, the parameters $r_0$ and $\rho_0$
are clearly correlated: the halos that are compact are systematically
denser. DM halos in all models are fairly consistent with the
observational points, except possibly for the CDM model that appears to form 
halos somewhat denser than observed. 
Note that the absence of halos at 
$r_0\lesssim 2h^{-1}$ kpc is due to our finite numerical resolution 
rather than the generic failure of these models to produce very 
compact halos. 
 The characteristic density of the DM halos
correlates strongly with halo mass in a way that reflects the mass 
dependence of the epoch of halo formation (NFW96): low-mass small 
halos collapse  at systematically higher redshifts (when the
universe was denser) and are therefore denser
than the larger higher-mass halos. Thus, the correlation observed in Figure 3
is likely to reflect the different formation epochs of individual
halos.

\pspicture(0.5,-1.5)(13.0,17.5)

\rput[tl]{0}(-1.5,17.9){\epsfxsize=12cm
\epsffile{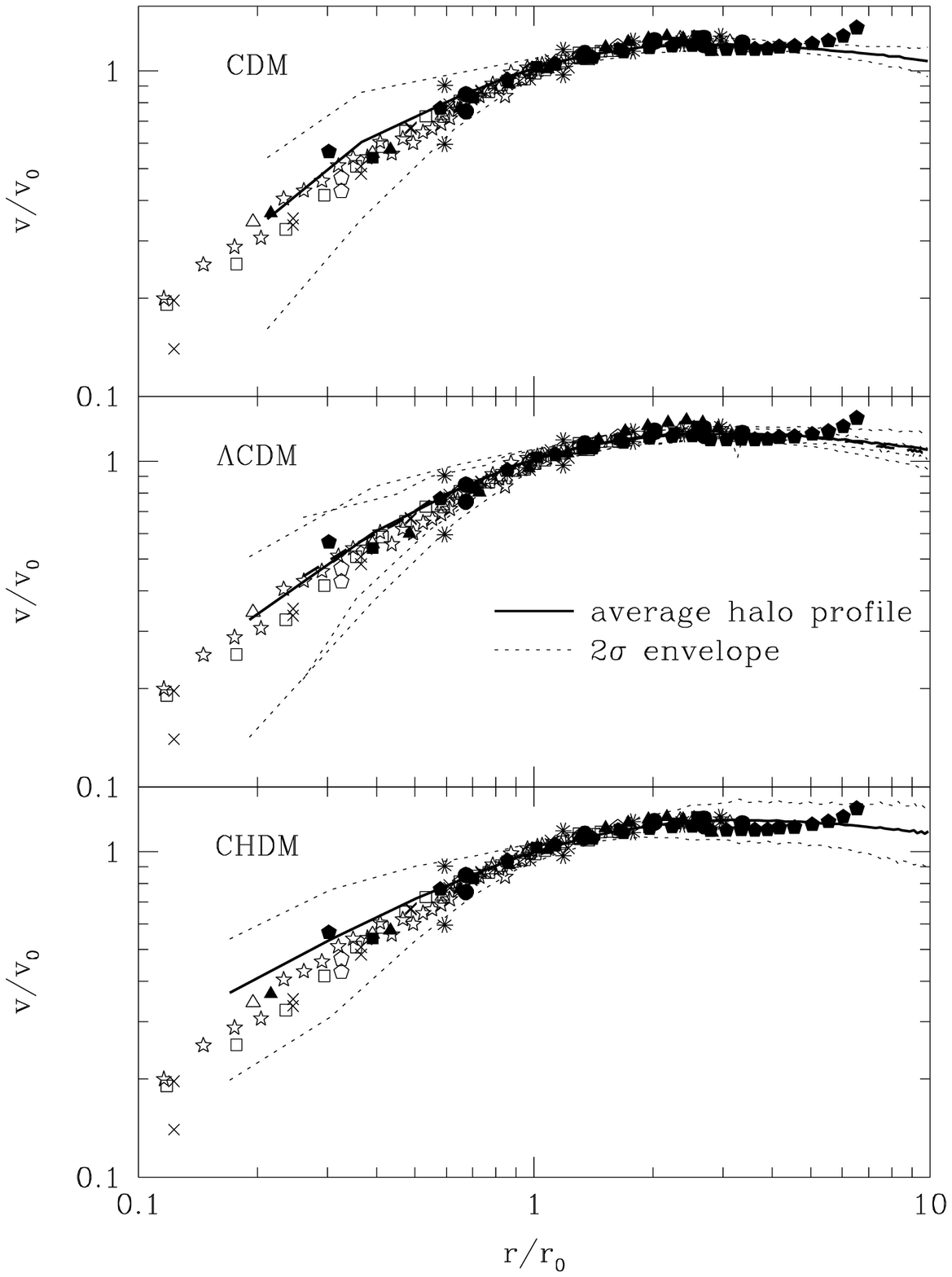}}

\rput[tl]{0}(0,6.0){
\begin{minipage}{8.7cm}
  \small\parindent=3.5mm {\sc Fig.}~2.--- Average normalized dark
  matter velocity profiles for halos formed in ({\em top panel}) CDM,
  ({\em middle panel}) $\Lambda$CDM, and ({\em bottom panel}) CHDM
  models with corresponding profiles of the dwarf galaxies from our
  sample.  The dotted lines show the $2\sigma$ envelope representing
  scatter of individual halo profiles around the average. It should be
  noted that although the velocity profiles of the hierarchically
  formed dark matter halos are on average consistent with the shape of
  observed rotation curves, the scatter in the inner regions of the
  halo velocity profiles is substantial.  This scatter possibly
  reflects real physical differences between individual halos.  The
  average profile from the larger-box \LCDMl simulation (with 2 times
  worse spatial and 8 times worse mass resolutions) is shown with a
  dashed line in the middle panel. This profile does not extend to
  values of $r/r_0$ which are as low as for the \LCDMs profile (due to
  worse spatial resolution). However, for values of $r/r_0$, where the
  two profiles overlap, they are indistinguishable.  This suggests that
  the shape is not affected by the finite size of the simulation box
  and mass resolution.  The peculiar upturn in the rotation curve of
  NGC2915 is discussed in \S 2.2.
\end{minipage}
}
\endpspicture

A similar correlation can be observed in the $r_{max}-v_{max}$
plane, shown in Figure 4 (values of $r_{max}$ and $v_{max}$ for each galaxy are given in Table 1).  The maximum point in a galaxy's DM velocity
profile and the corresponding radius is a nice set of 
physical parameters for comparison with simulations.  Ideally, such
a comparison would not force any pre-supposed fit to either the data or the
the simulated profiles.   Unfortunately,  most of the galaxy 
rotation curves in our sample do not extend to $r$ large enough to explicitly 
define the maximum velocity.  Therefore, we find $v_{max}$ by fitting the velocity
profile $V(r)$ in eq. (4) using parameters which produce a velocity
curve equivalent to the $\rho(r)$ fits discussed above:
 (a,b,g) = (1.5, 0.34, 0.9).  After each galaxy is fit with this profile,
we use the fit to determine its maxima.  We have, in a sense, tested
this procedure with a mock run with the simulated halos: we performed the
same fit to the inner profiles of the simulated halos (r$<30$h$^{-1}$kpc)
and then compared this to the maxima determined using a smooth (all
parameters free) fit {\em over all radii} of the halos. 
The values of $v_{max}$ found by these two methods are virtually identical.
Although there is some scatter between the corresponding values of $r_{max}$,
there is no systematic difference between the two. 
The rms difference in values of $r_{max}$ (determined by the two fitting 
procedures) is 
smaller than the scatter from halo to halo at any particular 
$v_{max}$.

\begin{figure*}[ht]
\pspicture(0,8.0)(18.8,22.5)

\rput[tl]{0}(-2.0,22.6){\epsfysize=13cm
\epsffile{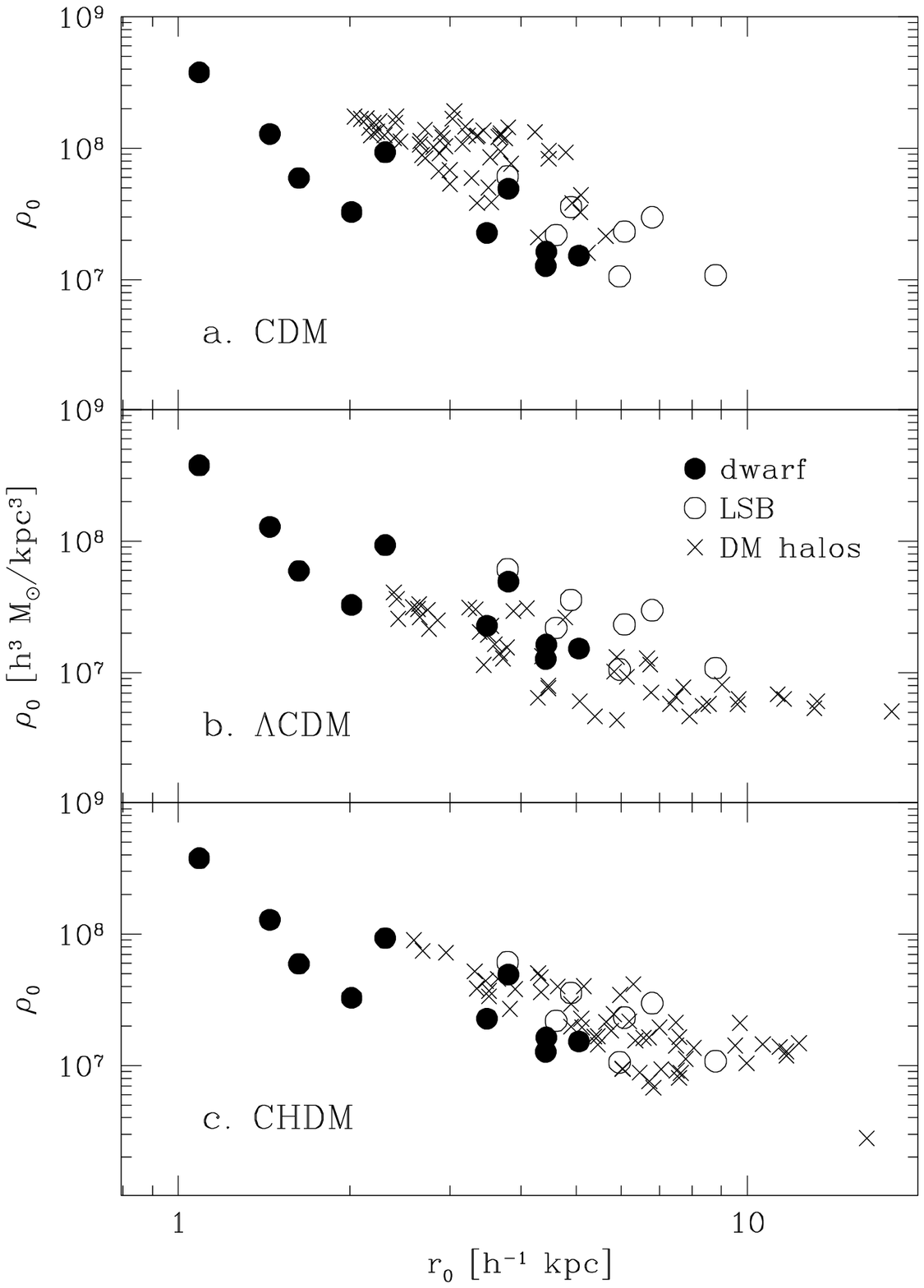}}

\rput[tl]{0}(7.5,22.6){\epsfysize=13cm
\epsffile{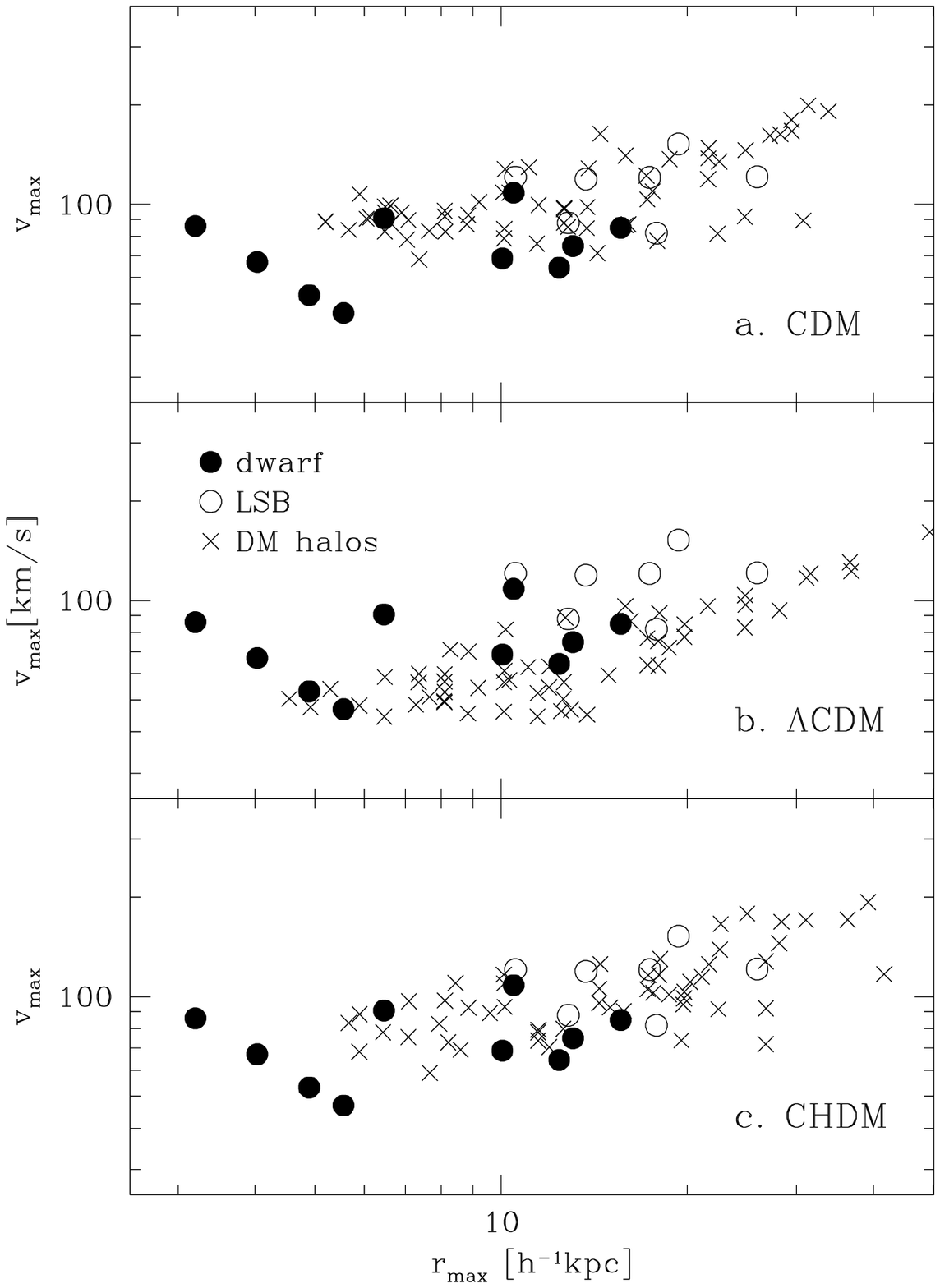}}

\rput[tl]{0}(-0.1,9.8){
\begin{minipage}{8.9cm}
  \small\parindent=3.5mm {\sc Fig.}~3.--- Correlation of the
  best-fit parameters $r_0$ and $\rho_0$ for the dwarf and LSB galaxies
  ({\em solid} and {\em open circles}, respectively) and for the dark matter
  halos ({\em crosses}) formed in (a) CDM, (b) $\Lambda$CDM, and (c) CHDM
  models. The correlation is consistent with the correlations of
  simulated DM halos: smaller halos are denser.
\end{minipage}
}

\rput[tl]{0}(9.6,9.8){
\begin{minipage}{8.9cm}
  \small\parindent=3.5mm {\sc Fig.}~4.--- Correlation of the
  maximum velocity of the rotation curves, $v_{max}$, and scale at
  which this maximum occurs, $r_{max}$, for observed galaxies and
  simulated DM halos (the symbols are the same as in Fig.3).
\end{minipage}
}

\endpspicture
\end{figure*}

We see from Figure 4 the trend that larger $r_{max}$ correspond
to larger $v_{max}$.  Such a trend was also seen in
NFW96 (their Fig.10). 
Note that when the $h$ factor is scaled out of the $r_{\rm max}$ 
axis, the three models lie very closely along each other
in this plot.  But since the observationally determined radii 
of the galaxies depend on $h$, the plots show the differences accordingly.

Note that we have fitted $\gamma=0.2$ from the observed galaxies rather
than a value derived from the simulations because of the larger 
scatter in the latter.  However, if we fix $\alpha=2$ and $\beta=3$ 
we can attempt to find the best-fitting ranges of $\gamma$ for
each of our simulated models.  The procedure
for finding best-fit gamma values for the halos is not entirely
straightforward since, depending on the
the mass of the halo (and intrinsic scatter), the characteristic
scale $r_{0}$, may be as small as the resolution limit, 
and the inner-slope will be poorly constrained.
That is, if we were to fix $\alpha=2$ and $\beta=3$ and fit the value
of $\gamma$ freely for each halo, 
the smaller halos ($M \lesssim 10^{11} M_{\odot}$)
would not offer any strong constraint,
and the fitting routine could yield spurious or unphysical values
for $\gamma$.  What we have done instead is compared the ``stacking'' of halos
for a range of $\gamma$ values (0-1.5).  We fit each halo
using logarithmic radial binning out to the virial radius, 
and only fit radii which were greater than twice the formal resolution.
This bin spacing emphasizes the inner profile shape. 
After fitting each halo for a particular value of $\gamma$ 
(with ($\alpha$, $\beta$) = (2,3))
we normalized each by $r_{0}$ and $\rho_{0}$ and effectively 
plotted them on top of each other.  The scatter in this stack is a 
nice way of determining how well the imposed profile shape 
fits a set of halos.  This procedure also allows us to treat large halos 
(which, within resolution, probe $r/r_{0} \ll 1$) and small halos 
equivalently.  Using this criteria, we find that the ranges
$\gamma_{\rm \Lambda CDM} \approx 0 - 0.4$, $\gamma_{\rm CDM}
\approx 0 - 0.5 $, and $\gamma_{\rm CHDM} \approx 0 - 0.7$ provide
the best fits.

%
\section{DISCUSSION}
%
The most important conclusion evident 
from the results presented in the previous
section is that there is {\it no statistically significant discrepancy 
between the shapes of rotation curves of simulated halos and rotation
curve shapes
of galaxies in our sample}. This conclusion is somewhat surprising
in light of the previous results (Flores \& Primack 94; Moore 1994;
NFW96) that indicated a significant discrepancy between numerical simulations
and rotation curve measurements. We see at least two possible 
explanations for the controversy. First, both Moore (1994) and
NFW neglected the fact that distances to these galaxies (and thus the physical 
scales of the rotation curves) 
are very poorly determined (with a typical error 
of $\sim 50$\% or more; L. van Zee, 
private communication).
With a distance uncertainty this large, it is hardly legitimate 
to make a raw comparison of the profiles at a specific physical distance
scale. The comparison between rotation curves
of dwarf galaxies and different analytic models was made by B95.
The comparison was made, however, in units of $r/r_b$ (see eq.(2)), 
in which case 
the uncertainty in distance in both $r$ and $r_b$ 
cancels out in the ratio (the same is true for our analysis that was done 
in units of $r/r_0$).
Indeed, the discrepancy between the analytic
model (1) and the data (Fig.1 in B95) was not as large as
was found by Moore (1994). Note that the study by
Flores \& Primack (1994) was also done using
dimensionless scale units: $r/b_{HI}$, 
 where $r$ is the physical scale of the 
rotation curves and $b_{HI}$ is the $HI$ disk scale length. The key point
is that the rotation curve {\it shape} is independent of
the uncertain distance to the galaxy.

The second possible source of discrepancy concerns the procedure
followed to compare the numerical results with observations.  
When comparing to the RC data, we
have used 
$(\alpha,\beta,\gamma)=(2,3,0.2)$  fit (motivated by the observed
galaxies' RCs; see Fig.1) only to determine $r_0$ and $v_0$ for the 
inner halo RCs (i.e.,
$r \leq 30h^{-1}$ kpc) so that these can be properly rescaled and 
compared to the dwarf RCs
in Fig. 2.  Other authors compare the data to a simple analytic 
fit to the entire halo profile (i.e., $r \lesssim
r_{vir}$), e.g. $\rho_{NFW}(r)$.  
 Therefore, when compared to the data, any
deviations of actual halo profiles from analytic fit were neglected. 
Although the universality of the mass
distribution in the DM halos is most likely real and reflects the 
self-similar nature of their formation, the associated scatter of 
the real profiles should not be neglected. Also,  
possible systematic deviations (especially in the inner, $r\lesssim
r_0$, regions) of the actual profiles from the analytic 
model (1) should be kept in mind. 
Our analysis shows that such deviations do, in fact, exist. 
Figure 5a shows the velocity profiles of a sample of the 
DM halos in our CDM simulation
normalized to their best fit values of the characteristic radius $r_s$
and rotational velocity at $r_s$. Figure 5b shows residuals between the 
halo velocity profiles and the analytic fit by the NFW profile.
All profiles are shown down to their spatial resolution. Figure 5b
shows that the halo rotational velocities at scales $r/r_s\lesssim 0.5$ 
are systematically lower than the rotational velocities
predicted by the best fit NFW profile.  
Fig. 5 shows that $\rho_{NFW}(r)$ is a rather good fit to our halos for $r
\gtrsim 0.03r_{vir}$, but not for the smaller scales that are relevant
to observed inner rotation curves.
Figure 6 in NFW96 and Figure 4 in NFW97 show
that similar deviations seem to exist in their simulations as well. 
It was suggested (J. Navarro, private communication) that 
the inner density distribution may depend on the dynamical state of the 
halo: the most relaxed halos may have systematically steeper inner
density profiles. We do not find such a trend for the halos analyzed 
in this paper. There does not seem to exist any correlation of the inner slope
of the density profiles or the 
concentration parameter, $c\equiv r_{vir}/r_s$,
with the dynamical state of the halo quantified by the fractional difference 
between the center of mass 
inside the halo virial radius and the halo center (the density peak):
$d_{CM}=\vert {\bf r}_{peak}-{\bf r}_{CM}\vert/r_{vir}$. However,
if a weak correlation does exist, it could be lost due to the rather 
large errors in determining $c$ ($\sim 30\%$) for small halos ($M_{vir}\sim
10^{11}-10^{12}h^{-1}$ M$_{\odot}$). 
Our analysis shows that 
although the whole ensemble of the halo profiles can be described 
reasonably well by a fixed set of the parameters $\alpha$, $\beta$, 
and $\gamma$, the scatter in these parameters among the individual
halos is substantial. Thus, for example, the average velocity profiles of 
halos shown in Figure 2 are very close to the observational points,
whereas the upper $2\sigma$-envelopes (the dotted lines) 
lie considerably higher in the inner regions of the rotation curves. 
The discrepancy between the analytic model of NFW and the data 
found by B95 is actually within $\lesssim 1\sigma$ from our 
average halo profiles. The scatter is not caused by poisson noise:
the halo rotation curves contain $\gtrsim 100$ particles inside the
innermost bin used in the fitting.  
There are different possible sources of this scatter. The set of
parameters ($\alpha$, $\beta$, $\gamma$)$=(2,3,0.2)$, used to make a
sensible comparison between data and models in the $r/r_0-v/v_0$ plane 
may not be the best description of the mass distribution in the simulated 
halos. However, our analysis shows that the true source is probably
real differences between mass distributions of different halos. For 
example, when we allow parameter $\alpha$ vary freely with other
parameters fixed to $\gamma =0.2$ and $\beta=3$, this results in a
wide range of best fit values for $\alpha$ ($\sim 0.7-2$). Also, the
density distribution in the inner regions  varies substantially from
halo to halo: the density profiles of some halos are considerably
steeper than profiles of the others. We can thus talk only about
``approximate universality'' of the profiles and some amount of scatter
will be introduced with any fixed set of parameters. 
Regardless of the 
nature of this scatter, it should not be neglected when making comparisons 
with the data. And it is this scatter which makes us  
conclude thus it is premature to claim a discrepancy between the mass 
distribution of hierarchically formed halos and  
observed rotation curves of dark matter dominated galaxies. 

\pspicture(0.0,-1.5)(13.0,18.0)

\rput[tl]{0}(-1.5,18.3){\epsfxsize=12cm
\epsffile{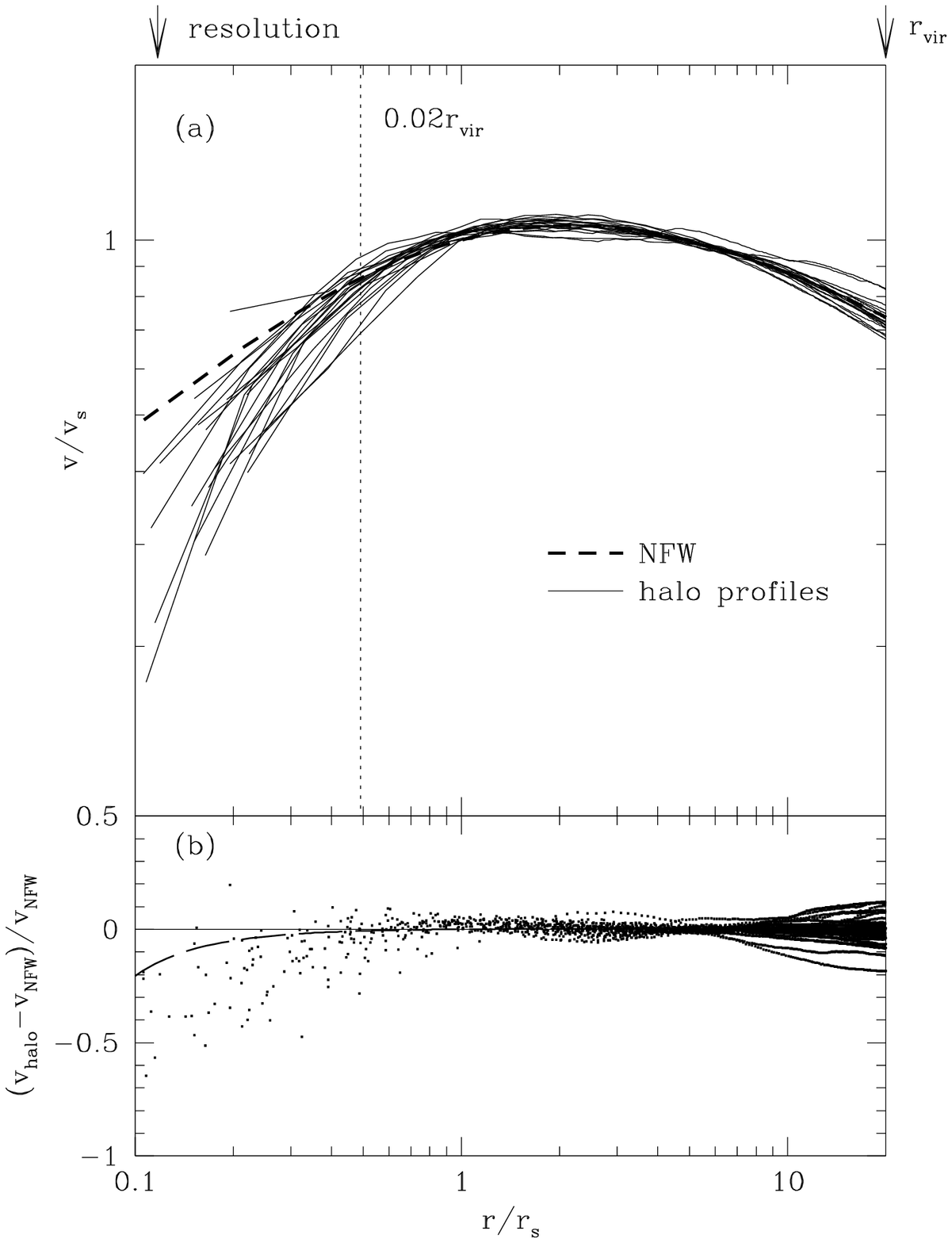}}

\rput[tl]{0}(-0.3,6.5){
\begin{minipage}{8.7cm}
\small\parindent=3.5mm
{\sc Fig.}~5.--- (a) Velocity profiles of the DM halos in our CDM simulation
normalized to their best fit values of the NFW profile:
the characteristic radius $r_s$ (see eq.[1])
and the rotational velocity at $r_s$ (the actual values of these
parameters for the halos used are in the range $r_s\sim 4-10h^{-1}$ kpc
and $v_s\sim 120-200h^{-1}$ km/s depending on the halo's virial mass). 
The dashed line shows the RC predicted by 
the NFW analytic density profile (eq. [1]). (b) Residuals between the 
halo velocity profiles and the analytic fit by the NFW profile.
All profiles are shown down to their spatial resolution. Note that
although the NFW profile provides a reasonably good fit at 
$0.5\lesssim r/r_s<20$ 
(corresponding approximately to $0.02\lesssim r/r_{vir}<1$),
 the rotational velocities of halos at scales $r/r_s\lesssim 0.5$ 
are systematically lower than the rotational velocities
predicted by the NFW profile. The long-dashed curve shows errors in 
rotation velocity due to the force softening, assuming the NFW profile 
(see \S 3.2). The errors due to softening are $\lesssim 5\%$ at 
the scales that were used in fitting ($x\gtrsim 0.2$, 
corresponding approximately to two resolution elements) and cannot 
account for $\sim 5-10$ times larger
deviations from the NFW profile observed at these scales. 
\end{minipage}
}
\endpspicture

Although it appears that we cannot reject any of the analyzed
cosmological models, we note that the cores of DM dominated galaxies
are potentially very useful probes of both the history of galaxy
formation and the underlying cosmological model. The inner regions of
DM halos are expected to be very sensitive to the merging history.  The
accretion of small dense satellites builds up a cuspy inner density
profile after only few mergers (Syer \& White 1997), even if the
initial density profile had a flat core. The slope $\gamma$ is expected
to be a function of the merger rate and the slope of the perturbation
spectrum on scales corresponding to the mass of a halo.  This result
can be used possibly to constrain models.  According to the analysis of
Syer \& White (1997), the steeper the slope of the power spectrum $n$
($P(k)\propto k^n$), the shallower the central density profile: $\gamma
\approx 3(3+n)/(5+n)$. If this result is qualitatively correct, the
shallower slope of $P(k)$ at cluster scales ($n\sim -2$), as compared
to the slope at galactic scales ($n\sim -2.5$), should result in
steeper central density profiles for cluster-size halos.  The rotation
curves of galaxies in our sample suggest a slope of $\gamma \sim
0.2-0.4$.  These values, according to the analysis of Syer \& White,
correspond to spectrum slopes of $n\approx -2.9$ and $n\approx -2.7$,
Note that $n_{\rm CDM}\approx-2.4$, $n_{\rm \Lambda CDM}\approx n_{\rm
  CHDM}\approx-2.6$ on the scales of the halos that we analyze.  The
slope $\gamma$ may potentially be a useful probe of the spectrum. Note
that our simulations show a considerable spread in the slopes of the
central density profiles. The physical processes which lead to
differences in the central density profiles seen in our halos (e.g.,
environment, dynamical state, etc.) are not clear and will be
investigated in a future study.

Navarro, Eke, \& Frenk (1996) suggested that the cores of dwarf
galaxies may also be sensitive to possible past violent star formation
bursts in these galaxies. They showed that a sudden loss of most of 
a galaxy's gas, blown away in the course of a star formation burst,
may lead to formation of a flat core, even if the initial density
distribution was coreless. However, the parameters used to model 
such an event were somewhat extreme, and it is not clear whether this 
mechanism would work for most of the dwarf galaxies. Moreover, a
significant fine tuning would be required to explain the observed 
degree of self-similarity of the mass distribution in the dwarf 
galaxies. 
In this paper
we showed that mass distribution in the LSB galaxies is, in fact, 
very similar to the mass distribution of the dwarf galaxies. This 
makes it even harder to explain such regularity with this mechanism 
alone, because many of the LSB galaxies are fairly massive systems
which are unlikely to have lost much of their gas in starbursts. 

Recently, Burkert \& Silk (1997) used an improved rotation curve 
measurement of the dwarf DDO154 to argue that discrepancy between 
the NFW profile and the observed mass distribution in this galaxy can be explained
by a dark spheroid of baryons with mass several times the mass of the
observed disk and comparable to the mass of the cold dark matter halo. 
This hypothesis must be tested using other galaxies and 
observations of the MACHOs in our galaxy that constitute
the main observational evidence for the existence of such massive 
baryonic halos. The observed self-similarity
of the mass distribution in the dark matter dominated galaxies which 
we analyze requires
that the distribution of the baryonic dark matter be self-similar
too and suggests a certain degree of ``conspiracy'' between baryonic 
DM and cold DM halos. The results presented in this paper show 
that the mass distribution in our hierarchically formed cold dark matter 
halos is consistent with the dynamics of the dark 
matter dominated galaxies and we think that it is somewhat premature to invoke 
an additional dark matter component. However, if further results on 
the distribution of MACHOs  shows that the mass contribution of these
objects is substantial, the possibility of such a component must be taken
into account. 
Note also that, as we discussed
in \S 2.2, the rotation curves of two of the dwarf galaxies, 
DDO154 
and NGC2915, show a peculiar behavior at large radii that cannot 
be described by any smooth model of density distribution. 
 
\section{CONCLUSIONS}
To summarize, our results are the following.
\begin{itemize}
\item Rotation curves of DM dominated
      dwarf and LSB galaxies have a similar shape. 
      This shape is inconsistent with $\rho_{NFW}(r)$
      (eq. [1]), 
      but is well fit by the coreless profile described by eq. (3)
      which has a shallower slope at small scales: 
      $\rho(r)\propto 1/r^{\gamma},{\ }\gamma\approx 0.2-0.4$,
corresponding to a steeply rising velocity curve $[v(r)\propto r^g,{\ }
g\approx 0.9-0.8]$.
\item We find that {\em on average} the velocity profiles of the halos formed in
      the hierarchical structure formation models analyzed in this paper 
       (CDM, $\Lambda$CDM,
      and CHDM) 
      and observed dark matter halos are in reasonably good agreement.
      We find a substantial amount of scatter in the central density profiles
      of individual halos around the average. The physical processes which lead 
      to differences in the central density profiles seen in our halos
      (e.g., environment, dynamical state, etc.) are not clear and will be
      the subject of a future study.  
\item The inner ($r<30 h^{-1}$ kpc) average density profiles of DM halos in 
      all of our simulations
      are well fit by model (3) with ($\alpha$, $\beta$, $\gamma$)$=$(2,3,0.2),
      and equivalently the rotation curves are well described by RC (4)
      with $(a,b,g)=(1.5,0.34,0.9)$.
      The profiles systematically deviate from the NFW profile (1) at small 
      scales. 
\item We find that dark matter dominated dwarf and LSB galaxies 
      show correlations between their characteristic density and 
      radius consistent with the correlations of hierarchically 
      formed DM halos: physically smaller halos are denser. 
      We find a similar correlation between the maximum of the rotation 
      curve, $v_{max}$, and the corresponding radius $r_{max}$. 
\end{itemize}
\acknowledgements
We would like to thank Erwin de Blok and Liese van Zee for sending us rotation 
curves in the electronic form. This work was supported by NASA and NSF
grants at NMSU and UCSC.  JSB acknowledges support from a GAANN predoctoral
fellowship at UCSC and thanks the Northern California Association of 
Phi Beta Kappa for a graduate scholarship.  Simulations reported here 
were done at NMSU, NCSA (at the University of Illinois), and Hebrew University.

\end{document}